\documentclass{elsart}

\usepackage{graphicx}
\usepackage{amsmath}
\usepackage{amsfonts}
\usepackage{amssymb}

\usepackage{color,epsfig}

\newcommand{\lsim}{\lower.7ex\hbox{$\;\stackrel{\textstyle<}{\sim}\;$}}

\begin{document}
\begin{frontmatter}

\title{Estimating the Explosion Time of Core-Collapse Supernovae \\
       from Their Optical Light Curves}
\author[HU,PENN]{D.F.~Cowen}
\author[HU,BONN]{A.~Franckowiak} 
\author[HU,BONN]{M.~Kowalski}
\address[HU]{Institut f\"ur Physik, Humboldt-Universit\"at zu Berlin, 12489 Berlin, Germany}
\address[PENN]{Departments of Physics and Astronomy and Astrophysics, Pennsylvania State University, State College, PA 16801 USA}

\address[BONN]{Physikalisches Institut, Universit\"at Bonn, 53115 Bonn}

\begin{abstract}

  Core-collapse supernovae are among the prime candidate sources of
  high energy neutrinos. Accordingly, the IceCube collaboration has
  started a program to search for such a signal. IceCube operates an
  online search for neutrino bursts, forwarding the directions of
  candidate events to a network of optical telescopes for immediate
  follow-up observations. If a supernova is identified from the
  optical observations, in addition to a directional coincidence a
  temporal $\gamma$-$\nu$ coincidence also needs to be established.
  To achieve this, we present a method for estimating the supernova
  explosion time from its light curve using a simple model. We test
  the model with supernova light curve data from SN1987A, SN2006aj and
  SN2008D and show that the explosion times can be determined with an
  accuracy of better than a few hours.

\end{abstract}

\end{frontmatter}

\section{Introduction}

Supernova explosions feature the interplay of all four known
fundamental forces.  A complete picture of supernova (SN) explosions
will therefore require true multi-messenger observations, with data
from traditional optical telescopes analyzed alongside coincident data
from neutrino and gravitational wave detectors.

To provide such multi-messenger data as well as to increase the
sensitivity to neutrinos from SNe, the IceCube
Collaboration~\cite{IceCube} together with the ROTSE
Collaboration~\cite{rotse} have set up an optical follow-up program
that triggers optical observations on multiplets of high-energy muon
neutrinos~\cite{Franck09} (a similar program has been inplemented by
the ANTARES Collaboration~\cite{ANTARES,D:2008hv}). A multiplet is
defined as at least two muon neutrinos from the same direction that
arrive within a short time window (e.g., $\sim 100$ s). When this
happens, an alert is issued to the four ROTSE-III telescopes, which
immediately observe the corresponding region in the sky. Successful
$\gamma$-$\nu$ coincident detection would allow one to infer the
existence of jets in SNe and would probe the expected gamma-ray
burst--supernova connection~\cite{Razzaque:2004yv,Ando:2005xi}.  

Absent any other corroborating astrophysical evidence, a standalone
neutrino doublet is not a physically interesting occurrence because
the background rate of such doublets from atmospheric muon neutrinos
in IceCube is $O$(10/yr).  However, given the very small number of SN
expected by random chance in the doublet's temporal and directional
windows, the significance of a coincident optical observation of a SN
rises dramatically: a neutrino doublet and an optical observation in a
coincidence time window typically assumed to be $\Delta t \sim 1$~day
(which we will show can be narrowed considerably) is of comparable
significance to the detection of a standalone neutrino
triplet~\cite{Kowalski:2007xb}.  (Neutrino triplets occur by chance
only once every few millennia and therefore their detection would be
intrinsically significant.)  A gain in sensitivity is thus achieved by
effectively lowering the neutrino multiplicity threshold (from $N=3$
to $N=2$) affording a factor of about two increase in the rate of
detectable SNe~\cite{Kowalski:2007xb}. In other words, the ability to 
narrow the coincidence time window provides a way
to reduce the level of accidental coincidences between neutrino
doublets and SNe.  This useful reduction can be achieved by rejecting
coincidences for which the neutrino doublet arrival time is
statistically too far outside the narrower time window obtained from
a fit to the SN explosion time.

This program thus relies on the ability to match the explosion time as
determined from the neutrino multiplet arrival time to that determined
from the optical data.  Smaller explosion time uncertainties result in 
better rejection of accidental coincidences, lending more significance to 
a coincidence detection. Previous studies have assumed that the explosion 
time can be known
with a precision of about one day without, however, supporting this
assumption with observational data (see,
e.g.,~\cite{Kowalski:2007xb,Ando:2005ka}).

In this paper we present the first study of the determination of the
explosion time, $t_0$, from the SN optical light curve.  We have
produced a generic model for the light curve that we test with
light curve data from SNe with known explosion times.  Such a study
became attractive in the last few years due to the recent fortuitous
discoveries of two nearby type Ib/c (stripped-core) SNe,
SN2008D~\cite{Soderberg:2008} and SN2006aj~\cite{2006Natur.442.1011P},
each with an associated X-ray flash presumably from the shock
breakout. The short X-ray flash provides a time stamp for the
explosion that can be compared to the one obtained from fitting the
optical light curve data. Furthermore, for obvious reasons the light
curve data for these SNe begin very early after the X-ray
flash, and as such are well-suited to the method described below,
because as with an X-ray flash, a neutrino trigger will enable early
optical observation of the target SN.  The only other SN
that has an explosion time known with even better precision is
SN1987A.  It is a low-luminosity type IIP SN with a light curve
very different from that of SN2008D and SN2006aj. Nevertheless, the
physics of the early part of the light curve is similar enough that we
find we can successfully extend our analysis to SN1987A as well.

The explosion time of SN1987A is taken to be the time of the MeV
neutrino burst.  For SN2008D and SN2006aj, we take the X-ray flash as
a rough proxy for the explosion time.  The latter is justified by
considering both radius and shock velocities for these SNe. The
estimated radius at which SN2008D's progenitor system becomes
optically transparent to X-rays, $r_*\lsim10^{12} {\rm
  cm}$~\cite{Soderberg:2008,Modjaz:2008}, is relatively small.  For
SN2006aj, a larger radius of $r_*\sim 5 \times 10^{12}$~cm is
estimated~\cite{Campana:2006,WMC:2008}, while for SN1987A a
photospheric radius of $r_*\sim 2 \times 10^{12}$~cm is
assumed~\cite{Ensman:1991td}. The maximum shock velocity at the shock
breakout has been computed as a function of radius, energy and mass
in~\cite{Matzner,chevalier08}. Inserting parameters for the SNe at
hand we obtain $\sim 0.5$~c for SN2008D and $\sim 0.1$~c for SN1987A.
The non-relativistic theory in~\cite{Matzner,chevalier08} yields a
maximum shock velocity for SN2006aj that exceeds the speed of
light. The authors in~\cite{WMC:2008} do a relativistic treatment and
estimate 0.85~c. One obtains the minimum time scale $t_{\rm min}
=r_*/v_s^{\rm max} = 70$~s for SN2008D, $t_{\rm min}= 200$~s for
SN2006aj and $t_{\rm min}=1300$~s SN1987A. While this crude
calculation underestimates by a factor of five the $6\times 10^3$~s
delay time between explosion and shock breakout predicted by a
detailed simulation of SN1987A \cite{Ensman:1991td}, it indicates that
for SN2008D and SN2006aj, the shock breakout is not expected to appear
much later than $5 \times t_{\rm min}\sim 10^3$~s after the explosion.
As will be shown in Sec.~\ref{sec:Results}, this theoretical time
scale for the shock propagation is much shorter then the resolution of
the fit on the time of explosion $t_0$ that we obtain for SN2008D. For
SN2006aj, it is comparable to the resolution of the light curve fits.

The remainder of this paper is organized as follows. In
Sec.~\ref{sec:Data} we present the light curve data and the model that
is used to analyze them. In Sec.~\ref{sec:Results} we present the
results of the light curve fits for SN2008D, SN2006aj and 
SN1987A. In Sec.~\ref{sec:early} we discuss the importance 
of early light curve data. We summarize the implications of our 
results in Sec.~\ref{sec:Conclusion}.

\section{Light Curve Data and Model} 
\label{sec:Data}

The SN2006aj and SN2008D light curves contain data from times
exceptionally soon after their putative explosions, making an accurate
estimation of SN explosion times feasible.  For SN2006aj we use
the U, B and V band data from the SWIFT UVOT~\cite{Campana:2006} and
for SN2008D we use the B, V, R and I band data from
FLWO~\cite{Modjaz:2008}.  Additional data from other telescopes is
available, but in order to avoid calibration problems arising from
different filter and instrument pass bands, we decided to work only
with data from a single source\footnote{Ref.~\cite{Mazzali:2008}
  provides a V band data point 4 hours after $t_0$. While we have not
  included it in the fits shown in this paper, we note that it fits
  the model prediction well.}.  For SN 1987A, we use the photometric
B, V, R, and I band data compiled and analyzed consistently by Hamuy
{\it et al.}~\cite{Hamuy}. The first data point is 1.14~days after the
explosion. Again, to avoid calibration problems, we do not use the
earlier discovery data points that exist for the V-band.

We estimate the explosion time by fitting light curves under the
assumptions of an initial blackbody emission from the rapidly cooling
shock breakout, followed by a phase dominated by the expansion of the
luminous shell. For the latter we test two distinct models.

{\bf Shock Breakout Phase:} To represent the shock breakout phase we
use the formulation from Waxman et al.\ \cite{WMC:2008}.  The flux
during the shock breakout phase of the SN light curve is
approximated by $\Phi_{BB} = I A$, where $A = 4\pi r^2$ is the area
and $I$ is the intensity.  The intensity is taken as proportional to
that produced by a blackbody at a fixed wavelength (we set $\lambda =
600$~nm, but this reference wavelength is not relevant for the results
presented here since it appears as multiplicative factor to the fitted
temperature). In addition to the explosion time $t_0$, the other free
parameters of the model are the radius and temperature at a fixed
reference time. Waxman et al.\ \cite{WMC:2008} give the SN radius $r
\propto \delta_t^{0.8}$ and the shock breakout temperature $T \propto
\delta_t^{-0.5}$, where $\delta_t = (t - t_0)$ is the elapsed time
since the explosion. Inserting these relations in the flux equation
yields:
\begin{equation}
   \Phi_{BB} = \frac{a_{1}}{\exp(a_{2}\delta_t^{0.5})-1}\delta_t^{1.6},
   \label{eq:PhiBB}
\end{equation}
with $a_{1}$, $a_{2}$ and $t_0$ free parameters.

{\bf Expansion Phase:} For the expansion phase we use either a simple
expanding photosphere model for the behavior of the light curve or the
more complex description from Arnett~\cite{Arnett:1982} that uses a
time-dependent diffusion equation.

\begin{figure}[htp]
   \centering
   \includegraphics[scale=0.5]{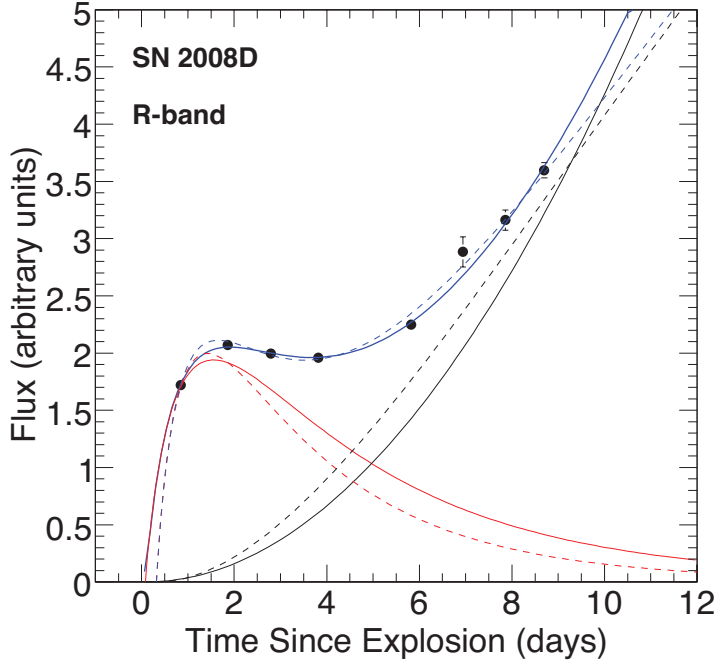}
   \caption{The rising part of the R-band light curve data for SN2008D
     from~\cite{Modjaz:2008} are shown along with the fit results. The
     fit model consists of a superposition of a blackbody spectrum
     (the initial ``bump" of the curve) and a model for the later
     emission (the rising part of the curve). The solid lines
     represent the fit results using the $t^2$ dependence for the late
     time emission; the dashed lines the Arnett model (the $t^2$
     formulation, which here gives $t_0 = 2\pm3$~hr, is preferred due
     to a better fit; see text in Sec.~\ref{sec:Results} for details).}
   \label{fig:SinglePlot}
   \vspace*{0.19cm}
\end{figure}

In the first model, the flux in the pure expansion phase is
approximated as
\begin{equation}
   \Phi_{t^2} = a_{3}\delta_t^{2},
   \label{eq:Phiat2}
\end{equation}
with $\delta_t$ defined above and $a_{3}$ and $t_0$ free parameters.
This $t^2$ assumption treats the SN photosphere as represented
by a blackbody of constant temperature, which expands with constant
velocity $v$~\cite{1998AAS,Riess1999}. The area of the photosphere, which is
directly proportional to the photon flux, then increases $\propto
(v\delta_t)^2$. This {\it ansatz} works remarkably well for the rising
part of type Ia SN light curves~\cite{Conley:2006tw}. The model has
one free parameter and, when combined with the blackbody emission
model, there are a total of four parameters in the fit to the light
curve.

As an alternative to the expanding photosphere model, we use the light
curve model of Arnett \cite{Arnett:1982} (also used
in~\cite{Soderberg:2008}), that assumes homologous expansion,
radiation pressure dominance, and $^{56}$Ni present in ejected matter
and distributed toward the center of the ejected mass. In this
alternative model there are two free parameters for the rising part of
the light curve model, so there are a total of five free parameters in
the fit to the light curve.

\begin{figure*}[htp]
   \centering
   \includegraphics[scale=0.7]{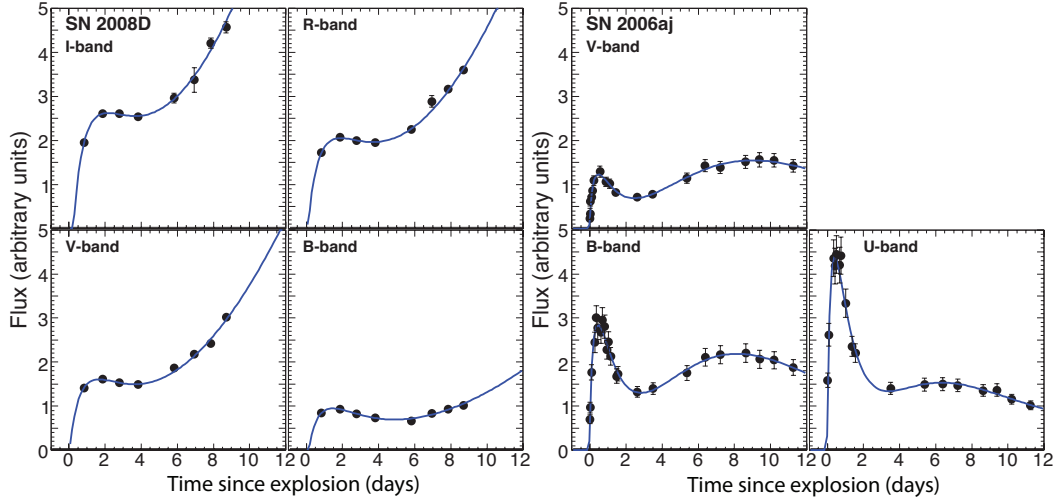}
   \caption{Early light curve data for type Ib/c SN2008D (left)
     from~\cite{Modjaz:2008} and for SN2006aj (right)
     from~\cite{Campana:2006} are shown and the fit performed for
     several optical bands.  For SN2008D the fit function is an
     initial blackbody spectrum followed by a $t^2$ dependence.  The
     same fit function has also been used for SN2006aj in the $t_0$
     analysis, but for illustrative purposes in this figure we show
     the fit using an initial blackbody spectrum followed by the
     Arnett formulation.  The fit result is shown as a solid line.}
   \label{fig:AllSevenPlots}
\end{figure*}

\begin{figure*}[htp]
   \centering
   \includegraphics[scale=0.8]{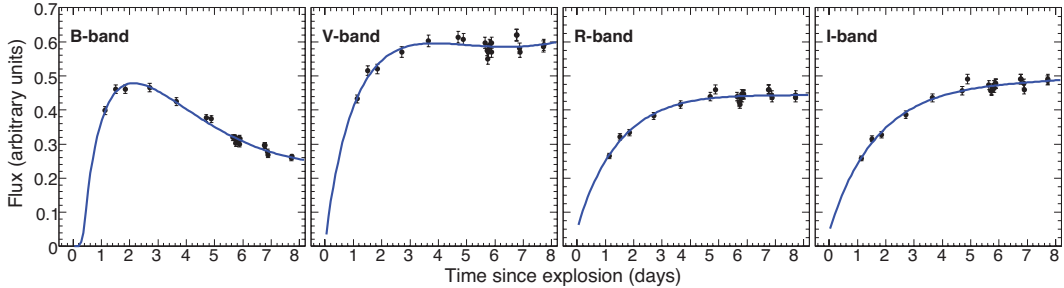}
   \caption{Early light curve data for the type II SN1987A from~\cite{Hamuy} 
     are shown.  For the fit function we use the initial blackbody
     spectrum followed by a $t^2$ dependence. }
   \label{fig:lc1987A}
\end{figure*}

As an example, Fig.~\ref{fig:SinglePlot} shows the results of the fits
to R-band light curve data of SN2008D. The full set of light curves
for SN2008D, SN2006aj and SN1987A are shown in Figs.~\ref{fig:AllSevenPlots}
and~\ref{fig:lc1987A}. A systematic evaluation of fits to all
available bands is the subject of the next section.

\section{Fit Results}
\label{sec:Results}

We fit our two models to the light curve data in multiple bands for
SN2006aj and SN2008D, as shown in Fig.~\ref{fig:AllSevenPlots}.  For
each fit we extract the initial explosion time, $t_0$, the error on
$t_0$, and the $\chi^2$ of the fit.

For SN2006aj, we find only marginal difference in the accuracy of the
fitted $t_0$ if we use the more complex Arnett formulation instead of
the simpler $t^2$.  (For the comparison, we restricted the fit to the
first six days, since the light curve of SN2006aj evolves faster than
other SNe, and for later times the $t^2$ approximation does not
hold.)  The agreement between both fit models is due to the fact that
the earliest part of the light curve is entirely dominated by emission
from the shock breakout and hence already strongly constrains
$t_0$. We obtain an average $t_0$ that is shifted by -0.04~days
relative to the X-ray flash, with a statistical error of about
0.005~days.

\begin{figure}[t!]
   \vspace*{0.3cm}
   \centering
   \includegraphics[scale=0.7]{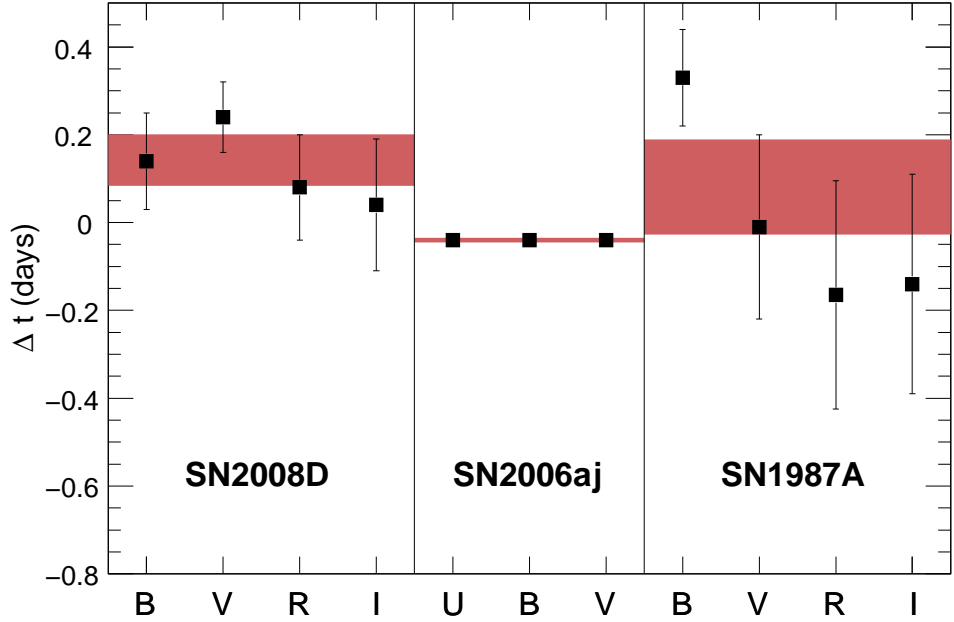}
   \caption{ Summary of the results of the fits to the light curves of
     SN2008D (left), SN2006aj (center) and SN1987A (right) in each
     optical band that was used.  The horizontal shaded regions are
     centered vertically on the error-weighted mean of $\Delta t$ (the
     difference between the fitted $t_0$ and the time of the X-ray
     flash or neutrino burst) and have a thickness corresponding to
     the error on the mean.  The $t^2$ formulation is used throughout
     since it provides comparable or better quality fits relative to
     that of Arnett.}
   \label{fig:SummaryPlot}
\end{figure}

The light curve data of SN2008D can also be fit by both the $t^2$ and
Arnett formulations.  However, the early data shown in
Figs.~\ref{fig:SinglePlot} and~\ref{fig:AllSevenPlots} is better
represented by the $t^2$ model, as determined by the quality of the
fit. We obtain from the fits a $\chi^2/{\rm NDF} = 15.9 / 16$ for the
sum of all four bands. Fitting with the Arnett formulation instead of
$t^2$ one obtains, with one additional fit parameter per band, a
$\chi^2$ that is significantly worse ($\chi^2/{\rm NDF} = 18.6/ 12$
for the sum of all data).  We hence proceed with the $t^2$ fit model 
as our default fit method. 
The fit results for all bands using the $t^2$ formulation
are shown in Fig.~\ref{fig:SummaryPlot}.

For SN2008D, whose light curves do not start so soon after the
explosion time, we find the fitted $t_0$ is consistent with zero for
three out of four bands (90\% CL), with an average error
of about 0.06~days. The largest outlier is the V-band, with
$t_0=0.24\pm0.08$~days. If the Arnett formulation is used instead of the 
default $t^2$ formulation, the estimated $t_0$ would be shifted by almost 5 
hours to late times (see also Fig.~\ref{fig:SinglePlot}) .

In contrast to SN2006aj and SN2008D, SN1987A has its date
of birth clearly marked by the observation of a short burst of
neutrinos.  Since its detection, SN1987A has been studied in great
depth, both observationally and theoretically. We cannot expect to
have such detailed information for future SNe unless they appear in
our own galaxy and hence, for the sake of simplicity, we adopt the
methodology already used above. We have fit the light curve data of
SN1987A with the model composed of the shock breakout according to
Eq.~\ref{eq:PhiBB} and the $t^2$ dependence for the expansion
phase. We fit the first eight days of data. Since the photometric
data~\cite{Hamuy} does not come with estimated uncertainties, we have
chosen them to be 0.03~mag to achieve $\chi^2/{\rm NDF} \approx 1$ in
the fits. The size of this assumed uncertainty roughly matches the
largest scatter of photometric data points observed during a single
night.  The fit results for four bands are shown in
Fig.~\ref{fig:SummaryPlot}.  In the figure, the larger error bar for
the V, R and I-band fits with respect to the B-band fits reflects the
fact that the shock breakout feature is not very evident for the
redder bands, as can be seen in Fig.~\ref{fig:lc1987A}.

We have explored whether we can improve the fits by incorporating some
key observations for SN1987A into the model, for example that an
almost constant bolometric luminosity was observed after the first day
after the explosion.  In our simple picture, this is achieved by
making the photosphere radius expand linearly with time, $r \propto
\delta_t$, while keeping the photosphere temperature dependence as
before: $T \propto \delta_t^{-0.5}$.  Reinserting this into
Eq.~\ref{eq:PhiBB} provides a slightly modified model for the shock
breakout phase. Fitting this shock breakout model results in a
systematic shift of $-0.3$ days for all bands. While the
B-band result is now consistent with the explosion time obtained from
the neutrino burst, the fits of the redder bands appear systematically
shifted.  Either way, we observe a deviation of the order of 0.3~days
for one of the bands indicating the size of the systematic uncertainty
involved in the extrapolation.  (Using Arnett's formulation for the
late times is not well justified for SN1987A.  Nevertheless, we note
that using it does not significantly worsen the fits or change the
conclusions.)

Summarizing, for SN1987A the light curve data starting 1.14~days after
the neutrino burst allows one to fit the explosion time with a fitting
error of about 0.2~days and a systematic error of about 0.3~days. The
systematic uncertainty reflects the crudeness of the light curve model
employed. Nevertheless, relative to simply using $t_0 = 1.14$~days,
the fitting technique yields a factor of $\sim 3$ improvement in the
$t_0$ measurement.

The fitted $t_0$ values demonstrate that an estimate of the explosion
time with an accuracy of much less than one day can be made using
simple analytic light curve models.  The estimates are robust on the
scale of a few hours across several independent optical bands.

\begin{figure}[t]
   \vspace*{0.4cm}
   \centering
   \includegraphics[scale=0.7]{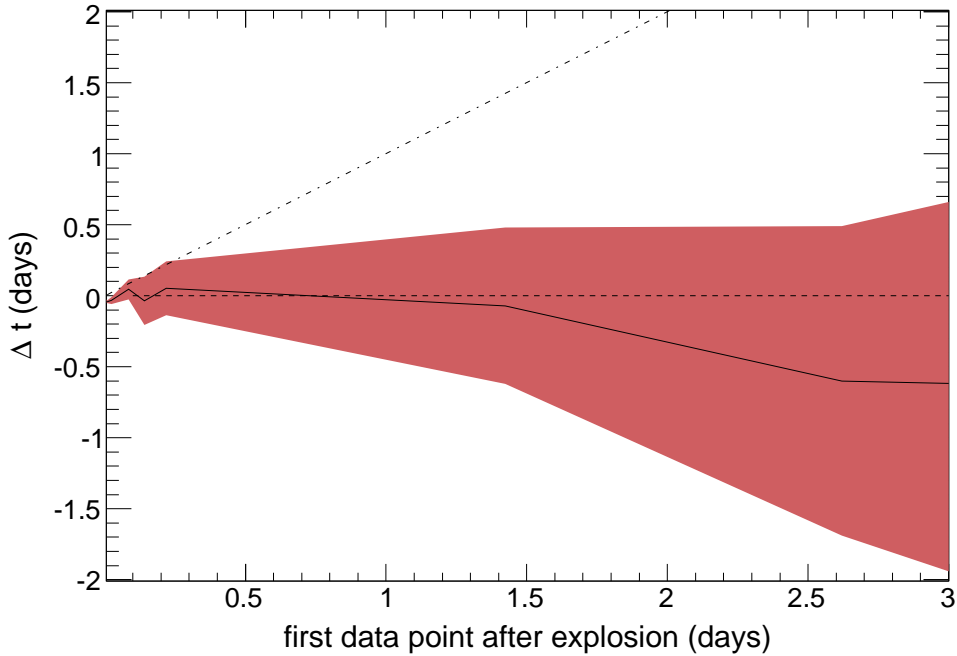}
   \caption{To quantify the importance of acquiring data points early
     in the SN light curve, we manually remove the  earliest V-band data
     points, one at a time, and refit the light curve for $t_0$ after
     each removal.  The solid black line shows the resulting fitted $t_0$
     values and the vertical height of the shaded region shows their
     1-$\sigma$ resolutions, as a function of the time of the earliest
     used data point in the fit.  The dot-dash line shows the value
     obtained for $t_0$ simply using the earliest available data point.
     Comparing this curve to the black line from the fitted $t_0$
     values, one sees that if there is a latency of roughly six or
     more hours after the putative explosion time before the first
     optical observation is made, the fitting technique provides the
     more accurate explosion time estimate.}

   \label{fig:RemovingPoints}
\end{figure}

\section{Importance of Early Light Curve Data}
\label{sec:early}

Using the SN2006aj data, we demonstrate the importance of the early
data points by manually removing the earliest data points, one at a
time, and re-fitting the data each time.  A summary of the result of
this exercise is shown in Fig.~\ref{fig:RemovingPoints}.  The figure
makes evident the importance of the early data points, showing how the
accuracy of the fitted $t_0$ depends strongly on these early data,
although the accuracy drops most dramatically after about a day. This
is consistent with the observation made for the other SNe: With a
first data point at $\sim 0.7$ days for SN2008D, the explosion date
can still be determined to within about 0.2~days, while for SN1987A,
with a first data point at 1.1 days after the explosion, the
uncertainty is around 0.3 days.  

It is also informative to compare the accuracy of the $t_0$ from the
fit with that obtained by simply using the first available point on
the light curve.  The dot-dash line in Fig.~\ref{fig:RemovingPoints} plot 
shows this $t_0$ estimate, which is simply the difference in time between 
the X-ray flash and the earliest remaining data point on the light curve. The
figure thus shows that if light curve data is acquired with about a
six hour or greater delay from the explosion time, the fitting
technique provides a more accurate and precise measure of the
explosion time than simply using the earliest point on the light
curve.

We have also studied the effectiveness of our method for estimating
the $t_0$ when the acquired light curve follows a multi-day cadence,
anticipating future survey telescopes such as
LSST~\cite{Ivezic:2008fe}.  Since we need to have an estimator of the
``true'' $t_0$ to do this, we took the SN2006aj light curve and kept
only a subset of its data, choosing those points separated by the LSST
cadence of about three days.  This was done for several distinct data
subsets.  In all cases, the fitted resolution on $t_0$ degraded
substantially, to about 1.5~days, in agreement with the results
shown in Fig.~\ref{fig:RemovingPoints} with about 2.5~days of early data
removed.  

Our method thus relies explicitly on the early detection of the light
curve.  Barring future serendipitous discoveries akin to SN2006aj and
SN2008D, we therefore rely on the neutrino-triggered optical follow-up
technique (mentioned earlier) to provide us with a light curve that
extends back suitably close in time to the actual $t_0$ of the
explosion.

\section{Conclusion}
\label{sec:Conclusion}

Both the IceCube and ANTARES Collaborations have started searching for
high energy neutrinos from SNe by implementing a
Target-of-Opportunity (ToO) program using robotic optical telescopes to
identify an optical counterpart to the neutrino
signal~\cite{Franck09,D:2008hv}.  A crucial ingredient therein is the
ability to determine the SN explosion time from the optically observed
light curve which allows one to establish a temporal coincidence with
the neutrino data.  Using a model for the early part of the light
curve, we show for the first time that one can estimate the explosion
times with an accuracy much better than the one day generally assumed
in the literature (see e.g.~\cite{Kowalski:2007xb,Ando:2005ka}).

We have fitted the light curves of three very different core-collapse
SNe: SN2008D, SN2006aj and SN1987A. For the expansion phase of the SN
we have tested two models, a simple $t^2$ model as well as a more
detailed model by Arnett.  We found that even with fewer parameters,
the $t^2$ dependence generally provids a better fit to the data as
well as a better match with the explosion time. We hence recommend
that this be used for a future coincidence search.

As shown in Fig.~\ref{fig:SummaryPlot}, the estimated $t_0$ and its
error, averaged over all available bands, is about $0.14\pm0.06 $~days
for SN2008D, $-0.04\pm0.005$~days for SN2006aj, and $0.08\pm0.11$ days
for SN1987A.  For SN2008D, the theoretical uncertainty associated with
the use of the time of the X-ray flash as the reference $t_0$ is
smaller than the resolutions on $t_0$ from our fits.  The fits in all
bands give explosion times that are slightly later than the time of
the X-ray flash, indicative of limitations in the rather simple
underlying physical model.  For SN2006aj, the explosion date was
determined from the fit to the light curve to be $3 \times 10^3$~s
before the X-ray flash. As mentioned earlier, this is larger than the
estimated time needed for the shock to propagate to the surface of the
progenitor. Resolving this discrepancy would require more detailed
modeling of the light curve and/or shock propagation.  In any case,
the discrepancy for both type Ib/c SNe investigated is $<4$~hrs, which
can be considered the characteristic size of the systematic
uncertainties in $t_0$.  Note that the resolutions on $t_0$ for both
SNe are longer than what is expected for the onset of gravitational
wave or high energy neutrino emission~\cite{Fryer:2003xt}.

The model does not take into account possible effects due to
circumstellar interactions, asymmetries in the ejecta or the
differences in the density profiles of the progenitors. These effects
might explain the observed deviations that are difficult to explain
with statistical errors alone.  Nevertheless, the fitted SN
explosion time $t_0$ represents a successful extrapolation of the data
to earlier times, and the magnitude of the extrapolation is large
compared to the quoted error.  This suggests that the model captures
dominant physical properties of the SN during the period
shortly after its explosion.

For the ongoing programs~\cite{Franck09,D:2008hv}, one can not expect
to have similarly detailed multiband, high signal-to-noise
observations as we had available for this study, hence we have focused
only on one band at a time. If multiband light curves are available,
one could do a combined fit and further improve the constraints.  In any
case, a future optical observation of a SN triggered by a
neutrino detector like IceCube or ANTARES should start early enough to
capture the initial shock breakout. If the initial shock breakout is
not observed, and the first observed point on the light curve is more
than 1-2~days after the actual explosion, the fits give large
uncertainties in the explosion time.  This illustrates the cardinal
importance of having fast follow-up capabilities in place to perform
ToO observations.

This work shows that the optical data can be fit accurately using the
formulation developed above, and that by doing so the statistical
significance of the coincidence can not only be quantified but also
significantly improved.

\section{Acknowledgements}

The authors would like to thank L. S. Finn, M. D. Kistler,
P. M\'esz\'aros and especially J. F. Beacom for helpful and
informative discussions.  DFC thanks the Deutscher Akademischer
Austausch Dienst (DAAD) Visiting Researcher Program and the Fulbright
Scholar Program, and acknowledges the support of the National Science
Foundation (NSF).  AF and MK acknowledge the support of the Deutsche
Forschungsgemeinschaft (DFG).

\end{document}